\documentclass[12pt,aps,showpacs,notitlepage]{revtex4-1}
\usepackage{amssymb,amsmath}
\usepackage{bm}
\usepackage[pdftex]{graphicx}
\usepackage{subfig}
\usepackage{color} 

\newcommand{\beq}{\begin{equation}}
\newcommand{\beqn}{\begin{eqnarray}}
\newcommand{\eeq}{\end{equation}}
\newcommand{\eeqn}{\end{eqnarray}}
\begin{document}

\title{Multifield Dynamics of Higgs Inflation}
\author{Ross N. Greenwood, David I. Kaiser, and Evangelos I. Sfakianakis}
\email{Email addresses: rossng@mit.edu; dikaiser@mit.edu; esfaki@mit.edu}
\affiliation{Center for Theoretical Physics and Department of Physics, \\
Massachusetts Institute of Technology, Cambridge, Massachusetts 02139 USA}
\date{\today}
\begin{abstract} Higgs inflation is a simple and elegant model in which early-universe inflation is driven by the Higgs sector of the Standard Model. The Higgs sector can support early-universe inflation if it has a large nonminimal coupling to the Ricci spacetime curvature scalar. At energies relevant to such an inflationary epoch, the Goldstone modes of the Higgs sector remain in the spectrum in renormalizable gauges, and hence their effects should be included in the model's dynamics. We analyze the multifield dynamics of Higgs inflation and find that the multifield effects damp out rapidly after the onset of inflation, because of the gauge symmetry among the scalar fields in this model. Predictions from Higgs inflation for observable quantities, such as the spectral index of the power spectrum of primordial perturbations, therefore revert to their familiar single-field form, in excellent agreement with recent measurements. The methods we develop here may be applied to any multifield model with nonminimal couplings in which the ${\cal N}$ fields obey an $SO ({\cal N})$ symmetry in field space.  
\end{abstract}
\pacs{04.62+v; 98.80.Cq. Published in {\it Physical Review D} 87 (2013): 064021}
\maketitle

\section{Introduction} 

The recent discovery at CERN of a scalar boson with Higgs-like properties \cite{Higgsdiscovery} heightens the question of whether the Standard Model Higgs sector could have played interesting roles in the early universe, at energies well above the electroweak symmetry-breaking scale. In particular, the suggestive evidence for the Higgs boson raises the possibility to return to an original motivation for cosmological inflation, namely, to realize a phase of early-universe acceleration driven by a scalar field that is part of a well-motivated model from high-energy particle physics \cite{LythRiotto,GuthKaiser,Mazumdar}.

Higgs inflation \cite{Higgsinfl} represents an elegant approach to building a workable inflationary model based on realistic ingredients from particle physics. In this model, a large nonminimal coupling of the Standard Model electroweak Higgs sector drives a phase of early-universe inflation. Such nonminimal couplings are generic: they arise as necessary renormalization counterterms for scalar fields in curved spacetime \cite{Fujii,Faraoni2004,BirrellDavies,Buchbinder}. Moreover, renormalization-group analyses indicate that for models with matter akin to the Standard Model, the nonminimal coupling, $\xi$, should grow without bound with increasing energy scale \cite{Buchbinder}. Previous analyses of Higgs inflation have found that $\xi$ typically grows by at least an order of magnitude between the electroweak symmetry-breaking scale and the inflationary scale \cite{SimoneHertzbergWilczek,BezrukovMass,Barvinsky}.

The Standard Model Higgs sector includes four scalar degrees of freedom: the (real) Higgs scalar and three Goldstone modes. In renormalizable gauges, all four scalar fields remain in the spectrum at high energies \cite{WeinbergQFT,Burgess,Barvinsky,Hertzberg,Mooij}. Thus the dynamics of Higgs inflation should be studied as a multifield model with nonminimal couplings. An important feature of multifield models, which is absent in single-field models, is that the fields' trajectories can turn within field space as the system evolves. Such turns are a necessary (but not sufficient) condition for multifield models to depart from the empirical predictions of simple single-field models \cite{GWBM,Groot,EastherGiblin,WandsReview,Langlois,PTGeometric,KMSBispectrum}.

In this paper we analyze the background dynamics of Higgs inflation, in which all four scalar fields of the Standard Model electroweak Higgs sector have nonminimal couplings. We find that multifield dynamics damp out quickly after the onset of inflation, before perturbations on cosmologically relevant length scales first cross the Hubble radius. As regards observable quantities like the power spectrum of primordial perturbations, the model therefore behaves effectively as a single-field model. The multifield dynamics remain subdominant in Higgs inflation because of the particular symmetries of the Higgs sector. Closely related models, which lack those symmetries, can produce conspicuous departures from the single-field case \cite{KMSBispectrum}.

We are principally interested here in the behavior of classical background fields and long-wavelength perturbations, which behave essentially classically. Therefore we bracket, for this analysis, the question of the unitarity of Higgs inflation. Conflicting conclusions have been advanced regarding whether the appropriate renormalization cut-off scale for this model should be $M_{\rm pl}$, $M_{\rm pl} / \sqrt{\xi}$, or $M_{\rm pl} / \xi$, where $M_{\rm pl} \equiv (8 \pi G)^{-1/2}$ is the reduced Planck mass \cite{Burgess,Espinosa,Barvinsky,Hertzberg,unitarity}. Even if Higgs inflation might  conclusively be shown to violate unitarity, the techniques developed here for the analysis of multifield dynamics will be relevant for related models that incorporate multiple scalar fields with nonminimal couplings and symmetries (such as gauge symmetries) that enforce specific relations among the couplings of the model. In particular, we expect that multifield effects in models with ${\cal N}$ scalar fields, in which the scalar fields obey an $SO ({\cal N})$ symmetry, should damp out rapidly.

In Section II, we briefly introduce the multifield formalism and establish notation. We apply the formalism to Higgs inflation in Section III, and in Section IV we analyze the behavior of the turn-rate, which quantifies the rate at which the background trajectory of the system deviates from a single-field case. We study how quickly the turn-rate damps to zero, both analytically and numerically, confirming that for Higgs inflation the turn-rate becomes negligible within a few efolds after the start of inflation. In Section V we turn to implications for observable features of the primordial power spectrum, confirming that multifield Higgs inflation reproduces the empirical predictions of previous single-field studies. Concluding remarks follow in Section VI.

\section{Multifield Dynamics}

Following the approach established in \cite{KMSBispectrum}, we consider models with ${\cal N}$ scalar fields in $(3 + 1)$ spacetime dimensions. We use Greek letters to label spacetime indices, $\mu , \nu = 0, 1, 2, 3$; lower-case Latin letters to label spatial indices, $i, j = 1, 2, 3$; and upper-case Latin letters to label field-space indices, $I, J = 1, 2, ... , {\cal N}$. We also work in terms of the reduced Planck mass, $M_{\rm pl} \equiv (8 \pi G )^{-1/2}$. In the Jordan frame, the action takes the form
\beq
S_{\rm Jordan} = \int d^4 x \sqrt{ - \tilde{g} } \left[ f (\phi^I ) \tilde{R} - \frac{1}{2} \delta_{IJ} \tilde{g}^{\mu\nu} \partial_\mu \phi^I \partial_\nu \phi^J - \tilde{V} (\phi^I ) \right] .
\label{SJordan}
\eeq
Here $f (\phi^I)$ is the nonminimal coupling function, and we use tildes for quantities in the Jordan frame. We perform a conformal transformation to the Einstein frame by rescaling the spacetime metric tensor,
\beq
g_{\mu\nu} (x) = \frac{2}{ M_{\rm pl}^2} f (\phi^I (x) ) \> \tilde{g}_{\mu\nu} (x) ,
\label{conftrans}
\eeq
so that the action in the Einstein frame becomes \cite{DKconf}
\beq
S_{\rm Einstein} = \int d^4 x \sqrt{- g} \left[ \frac{M_{\rm pl}^2}{2} R - \frac{1}{2} {\cal G}_{IJ} g^{\mu\nu} \partial_\mu \phi^I \partial_\nu \phi^J - V (\phi^I ) \right] .
\label{SEinstein}
\eeq
The potential in the Einstein frame, $V$, is related to the Jordan-frame potential, $\tilde{V}$, as
\beq
V (\phi^I ) = \frac{M_{\rm pl}^4}{4 f^2 (\phi^I )} \tilde{V} (\phi^I ) ,
\label{VE}
\eeq
and the coefficients of the noncanonical kinetic terms are \cite{SalopekBondBardeen,DKconf}
\beq
{\cal G}_{IJ} (\phi^K) = \frac{M_{\rm pl}^2}{2 f (\phi^I )} \left[ \delta_{IJ} + \frac{3}{ f(\phi^I )} f_{, I } f_{, J} \right] ,
\label{GIJ}
\eeq
where $f_{, I} = \partial f / \partial \phi^I$. The nonminimal couplings induce a field-space manifold in the Einstein frame that is not conformal to flat; ${\cal G}_{IJ}$ serves as a metric on the curved manifold \cite{DKconf}. Therefore we adopt the covariant approach of \cite{KMSBispectrum}, which respects the curvature of the field-space manifold.

Varying Eq. (\ref{SEinstein}) with respect to $\phi^I$ yields the equation of motion,
\beq
\Box \phi^I + g^{\mu\nu} \Gamma^I_{\> JK} \partial_\mu \phi^J \partial_\nu \phi^K - {\cal G}^{IK} V_{, K} = 0 ,
\label{phieom}
\eeq
where $\Box \phi^I \equiv g^{\mu\nu} \phi^I_{\>\> ; \mu ; \nu}$ and $\Gamma^I_{\>\> JK} (\phi^L)$ is the Christoffel symbold for the field-space manifold, calculated in terms of ${\cal G}_{IJ}$. We expand each scalar field to first order around its classical background value,
\beq
\phi^I (x^\mu) = \varphi^I (t) + \delta \phi^I (x^\mu) ,
\label{phivarphi}
\eeq
and also expand the scalar degrees of freedom of the spacetime metric to first order around a spatially flat Friedmann-Robertson-Walker (FRW) metric \cite{MFB,BTW,MalikWands}
\beq
ds^2 = - (1 + 2 A) dt^2 + 2 a ( \partial_i B) dx^i dt + a^2 \left[ (1 - 2 \psi) \delta_{ij} + 2 \partial_i \partial_j E \right] dx^i dx^j ,
\label{ds2}
\eeq
where $a (t)$ is the scale factor. We further introduce a covariant derivative with respect to the field-space metric and a directional derivative along the background fields' trajectory, such that for any vector $A^I$ in the field-space manifold we have
\beq
\begin{split}
{\cal D}_J A^I &= \partial_J A^I + \Gamma^I_{\>\> JK} A^K , \\
{\cal D}_t A^I &\equiv \dot{\varphi}^J {\cal D}_J A^I = \dot{A}^I + \Gamma^I_{\>\> JK} A^J \dot{\varphi}^K ,
\end{split}
\label{covderivs}
\eeq
where overdots denote derivatives with respect to cosmic time, $t$. 

To background order, Eq. (\ref{phieom}) becomes
\beq
{\cal D}_t \dot{\varphi}^I + 3 H \dot{\varphi}^I + {\cal G}^{IK} V_{, K} = 0 ,
\label{eomvarphi}
\eeq
where all quantities involving ${\cal G}_{IJ} (\phi^K)$, $V (\phi^I)$, and their derivatives are evaluated at background order in the fields: ${\cal G}_{IJ} \rightarrow {\cal G}_{IJ} (\varphi^K )$ and $V \rightarrow V (\varphi^I)$. Following \cite{GWBM} we distinguish between adiabatic and entropic directions in field space by introducing a unit vector
\beq
\hat{\sigma}^I \equiv \frac{\dot{\varphi}^I}{\dot{\sigma}} ,
\label{hatsigma}
\eeq
where 
\beq
\dot{\sigma} \equiv \vert \dot{\varphi}^I \vert = \sqrt{ {\cal G}_{IJ} \dot{\varphi}^I \dot{\varphi}^J } .
\label{dotsigma}
\eeq
The operator
\beq
\hat{s}^{IJ} \equiv {\cal G}^{IJ} - \hat{\sigma}^I \hat{\sigma}^J
\label{sIJ}
\eeq
projects onto the subspace orthogonal to $\hat{\sigma}^I$. Eq. (\ref{eomvarphi}) then simplifies to
\beq
\ddot{\sigma} + 3 H \dot{\sigma} + V_{, \sigma} = 0 
\label{eomsigma}
\eeq
where 
\beq
V_{, \sigma} \equiv \hat{\sigma}^I V_{, I} .
\label{Vsigma}
\eeq
The background dynamics likewise take the simple form
\beq
\begin{split}
H^2 &= \frac{1}{ 3 M_{\rm pl}^2} \left[ \frac{1}{2} \dot{\sigma}^2 + V \right] , \\
\dot{H} &= - \frac{1}{ 2 M_{\rm pl}^2} \dot{\sigma}^2 ,
\end{split}
\label{Friedmannsigma}
\eeq
where $H \equiv \dot{a} / a$ is the Hubble parameter.

We may also separate the perturbations into adiabatic and entropic directions. Working to first order in perturbations, we introduce the gauge-invariant Mukhanov-Sasaki variables \cite{MFB,BTW,MalikWands,Footnote1}
\beq
Q^I \equiv \delta \phi^I + \frac{\dot{\varphi}^I}{H} \psi 
\label{Q}
\eeq
and the projections
\beq
\begin{split}
Q_\sigma &\equiv \hat{\sigma}_I Q^I , \\
\delta s^I &\equiv \hat{s}^I_{\>\> J} Q^J .
\end{split}
\label{Qsigmadeltas}
\eeq
The gauge-invariant curvature perturbation may be defined as ${\cal R}_c \equiv \psi - [H / (\rho + p)] \delta q$, where the perturbed energy-momentum flux is given by $T^0_{\>\> i} = \partial_i \delta q$ \cite{BTW,MalikWands}. We then find that ${\cal R}_c$ is proportional to $Q_\sigma$ \cite{KMSBispectrum}
\beq
{\cal R}_c = \frac{H}{\dot{\sigma}} Q_\sigma .
\label{RcQsigma}
\eeq
Expanding Eq. (\ref{phieom}) to first order and using the projections of Eq. (\ref{Qsigmadeltas}), the perturbations $Q_\sigma$ and $\delta s^I$ obey \cite{KMSBispectrum}
\beq
\begin{split}
\ddot{Q}_\sigma + 3 H \dot{Q}_\sigma &+ \left[ \frac{k^2}{a^2} + {\cal M}_{\sigma\sigma} - \omega^2 - \frac{1}{M_{\rm pl}^2 a^3} \frac{d}{dt} \left( \frac{a^3 \dot{\sigma}^2}{H} \right) \right] Q_\sigma \\
&\quad\quad\quad = 2 \frac{d}{dt} \left( \omega_J \delta s^J \right) - 2 \left( \frac{ V_{, \sigma}}{\dot{\sigma}} + \frac{\dot{H}}{H} \right) \left( \omega_J \delta s^J \right) 
\end{split}
\label{eomQsigma}
\eeq
and
\beq
\begin{split}
{\cal D}_t^2 \delta s^I + \left[ 3 H \delta^I_{\>\> J} + 2 \hat{\sigma}^I \omega_J \right] {\cal D}_t \delta s^I &+ \left[ \frac{k^2}{a^2} \delta^I_{\>\> J} + {\cal M}^I_{\>\> J} - 2 \hat{\sigma}^I \left( {\cal M}_{\sigma J} + \frac{\ddot{\sigma}}{\dot{\sigma}} \omega_J \right) \right] \delta s^J \\
&\quad\quad = - 2 \omega^I \left[ \dot{Q}_\sigma + \frac{\dot{H}}{H} Q_\sigma - \frac{\ddot{\sigma}}{\dot{\sigma}} Q_\sigma \right] ,
\end{split}
\label{eomdeltas}
\eeq
where the mass-squared matrix is 
\beq
\begin{split}
{\cal M}^I_{\>\> J} &\equiv {\cal G}^{IK} \left( {\cal D}_J {\cal D}_K V \right) - {\cal R}^I_{\>\> LMJ} \dot{\varphi}^L \dot{\varphi}^M , \\
{\cal M}_{\sigma J} &\equiv \hat{\sigma}_I {\cal M}^I_{\>\> J} , \>\> {\cal M}_{\sigma\sigma} \equiv \hat{\sigma}_I \hat{\sigma}^J {\cal M}^I_{\>\> J} .
\end{split}
\eeq
The turn-rate \cite{PTGeometric,KMSBispectrum} is given by
\beq
\omega^I \equiv {\cal D}_t \hat{\sigma}^I = - \frac{1}{\dot{\sigma}} V_{, K} \hat{s}^{IK} ,
\label{omegadef}
\eeq
and $\omega \equiv \vert \omega^I \vert$. Eqs. (\ref{eomQsigma}) and (\ref{eomdeltas}) decouple if the turn-rate vanishes, $\omega^I = 0$. In that case, $Q_\sigma$ evolves just as in the single-field case \cite{MFB,BTW,MalikWands,PTGeometric,KMSBispectrum}. Given Eq. (\ref{RcQsigma}), that means that the power spectrum of primordial perturbations, ${\cal P}_{\cal R}$, would also evolve as in single-field models. Thus a necessary (but not sufficient) condition for multifield models of this form to deviate from the empirical predictions of simple single-field models is for the turn-rate to be nonnegligible for some duration of the fields' evolution, $\omega^I \neq 0$.

\section{Application to Higgs Inflation}

The matter contribution to Higgs inflation \cite{Higgsinfl} consists of the Standard Model electroweak Higgs sector, which may be written as a doublet of complex scalar fields,
\beq
h = \left( \begin{array}{c} h^+ \\ h^0 \end{array} \right) .
\label{SMHiggs}
\eeq
The complex fields $h^+$ and $h^0$ may be further decomposed into (real) scalar degrees of freedom,
\beq
\begin{split}
h^+ &= \frac{1}{\sqrt{2}} \left( \chi^1 + i \chi^2 \right) , \\
h^0 &= \frac{1}{\sqrt{2}} \left( \phi + i \chi^3 \right) ,
\end{split}
\label{phi+phi0}
\eeq
where $\phi$ is the Higgs scalar and $\chi^a$ (with $a = 1, 2, 3$) are the Goldstone modes. In the Jordan frame, the potential $\tilde{V} (\phi^I)$ depends only on the combination
\beq
h^\dagger h = \frac{1}{2} \left[ \phi^2 + {\bm \chi}^2 \right] ,
\label{hdaggerh}
\eeq
where ${\bm \chi} = (\chi^1, \chi^2,\chi^3)$ is a 3-vector of the Goldstone fields. In particular, the symmetry-breaking potential may be written
\beq
\tilde{V} (\phi^I) = \frac{\lambda}{4} \left( \phi^2 + {\bm \chi}^2 - v^2 \right)^2 ,
\label{V1}
\eeq
in terms of the vacuum expectation value, $v$. For the Standard Model, $v = 246$ GeV $\ll M_{\rm pl}$. For Higgs inflation, the nonminimal coupling function is given by
\beq
f (\phi^I ) = \frac{M_0^2}{2}  + \xi h^\dagger h = \frac{1}{2} \left[ M_0^2 + \xi \left(\phi^2 + {\bm \chi}^2 \right) \right] ,
\label{f}
\eeq
where $M_0^2 \equiv M_{\rm pl}^2 - \xi v^2$ and $\xi > 0$ is the dimensionless nonminimal coupling constant. In Higgs inflation, we take $\xi \sim {\cal O} (10^4)$ \cite{Higgsinfl}, and therefore we may safely set $M_0^2 = M_{\rm pl}^2$. In the Einstein frame, the potential gets stretched by the nonminimcal coupling function $f (\phi^I)$ according to Eq. (\ref{VE}). Given Eqs. (\ref{V1}) and (\ref{f}), this yields
\beq
V (\phi^I) = \frac{\lambda M_{\rm pl}^4 \left( \phi^2 + {\bm \chi}^2 - v^2 \right)^2}{ 4 \left[ M_{\rm pl}^2 + \xi \left( \phi^2 + {\bm \chi}^2 \right) \right]^2}  .
\label{VE2}
\eeq
The model is thus symmetric under rotations among $\phi$ and $\chi^a$ that preserve the magnitude $\sqrt{ \phi^2 + {\bm \chi}^2 }$. When written in the ``Cartesian" field-space basis of Eq. (\ref{phi+phi0}), in other words, the $SU (2)$ electroweak gauge symmetry manifests as an $SO (4)$ spherical symmetry in field space. 

For any model with ${\cal N}$ real-valued scalar fields that respects an $SO ({\cal N})$ symmetry, the background dynamics depend on just three initial conditions: the initial magnitude and initial velocity along the radial direction in field space, and the initial velocity perpendicular to the radial direction. Without loss of generality, therefore, we may analyze the background dynamics of Higgs inflation in terms of just two real-valued scalar fields, $\phi$ and $\chi$, and we may set $\chi (0) = 0$, specifying only initial values for $\phi (0)$, $\dot{\phi} (0)$, and $\dot{\chi} (0)$. This reduction in the effective number of degrees of freedom stems entirely from the gauge symmetry of the Standard Model electroweak sector. The remaining dependence on $\dot{\chi}$, meanwhile, distinguishes the background dynamics from a genuinely single-field model, in which one neglects the Goldstone fields altogether. For the remainder of this paper, we exploit the gauge symmetry to consider only a single Goldstone mode, ${\bm \chi} \rightarrow \chi$, reducing the problem to that of a two-field model. Then $f (\phi^I)$ and $V (\phi^I)$ depend on the background fields only in the combination
\beq
r \equiv \sqrt{ \phi^2 + \chi^2 } .
\label{r}
\eeq

\begin{figure}
\centering
\includegraphics[width=4in]{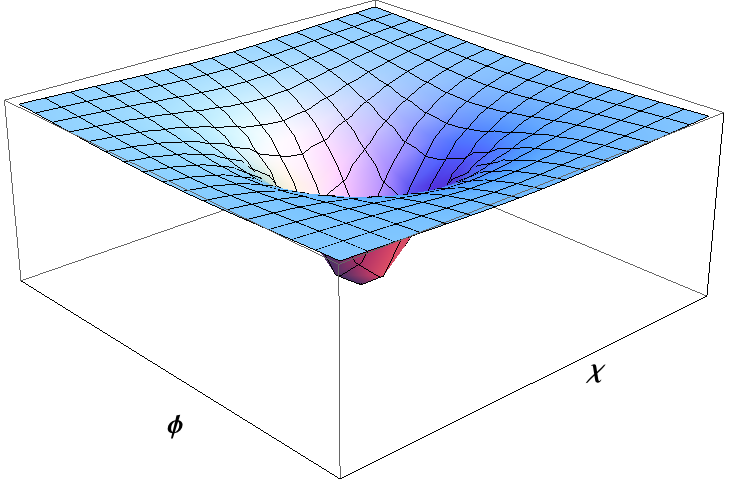}
\caption{\small \baselineskip 14pt The potential for Higgs inflation in the Einstein frame, $V (\phi, \chi)$. Note the flattening of the potential for large field values, which is quite distinct from the behavior of the Jordan-frame potential, $\tilde{V} (\phi, \chi)$ in Eq. (\ref{V1}).}
\label{HiggsPotential}
\end{figure}

Previous analyses \cite{Higgsinfl,SalopekBondBardeen,FakirUnruh,MakinoSasaki,DKnperts} which considered single-field versions of this model (neglecting the Goldstone modes) found successful inflation for field values $\xi \phi^2 \gg M_{\rm pl}^2$. We confirm this below for the multifield case including the Goldstone modes. The reason is easy to see from Eq. (\ref{VE2}). In the limit $\xi (\phi^2 + \chi^2 ) = \xi r^2 \gg M_{\rm pl}^2$, the potential in the Einstein frame becomes very flat, approaching
\beq
V (\phi^I ) \rightarrow \frac{\lambda M_{\rm pl}^4}{4 \xi^2} \left[ 1 + {\cal O} \left( \frac{M_{\rm pl}^2}{\xi r^2 } \right) \right].
\label{VEflat}
\eeq
See Fig. \ref{HiggsPotential}. 

Given $\xi \sim 10^4$, the initial energy density for this model lies well below the Planck scale, $\rho \simeq V \simeq \lambda M_{\rm pl}^4 / \xi^2 \sim 10^{-9} M_{\rm pl}^4$. In fact, as we will see, successful slow-roll inflation (producing at least 70 efolds of inflation) occurs for initial values of the fields below the Planck scale, unlike in models of chaotic inflation with polynomial potentials that lack nonminimal couplings. Moreover, as emphasized in \cite{Higgsinfl}, the flattening of the potential in the Einstein frame at large field values makes Higgs inflation easily compatible with the latest observations of the spectral index, $n_s$. Ordinary chaotic inflation with a $\lambda \phi^4$ potential and minimal coupling, on the other hand, yields a spectral index outside the 95\% confidence interval for the best-fit value of $n_s$ \cite{KomatsuWMAP,WMAP9}. Below we confirm this behavior for Higgs inflation even when the Goldstone degrees of freedom are included.

The field-space metric ${\cal G}_{IJ}$ is determined by the nonminimal coupling function, $f (\phi^I)$, and its derivatives. Explicit expressions for the components of ${\cal G}_{IJ}$ for a two-field model with arbitrary couplings, $\xi_\phi$ and $\xi_\chi$, are given in the Appendix of \cite{KMSBispectrum}. In the case of Higgs inflation, the $SU(2)$ gauge symmetry enforces $\xi_\phi = \xi_\chi = \xi$. Given this symmetry, the convenient combination, $C (\phi^I)$, introduced in the Appendix of \cite{KMSBispectrum} becomes
\beq
C (\phi^I ) = 2f +  6 \xi^2 \left( \phi^2 + \chi^2 \right) = M_{\rm pl}^2 + \xi (1 + 6 \xi) r^2 .
\label{C}
\eeq
For $\xi_\phi = \xi_\chi = \xi$, the Ricci curvature scalar for the field-space manifold, as calculated in \cite{KMSBispectrum}, takes the form
\beq
{\cal R} = \frac{4\xi}{C^2} \left[ C + 3\xi M_{\rm pl}^2  \right] .
\label{Ricci1}
\eeq
During inflation, when $\xi r^2 \gg M_{\rm pl}^2$, this reduces to
\beq
{\cal R} \rightarrow \frac{2}{3\xi r^2}  \ll M_{\rm pl}^{-2} ,
\label{Ricci2}
\eeq
indicating that the field-space manifold has a spherical symmetry with radius of curvature $r_c \sim \sqrt{ \xi} \> r $. As shown in \cite{KMSBispectrum}, the curvature of the field-space manifold remains negligible in such models until the fields satisfy $\xi r^2 \ll M_{\rm pl}^2$, near the end of inflation.

From Eq. (\ref{eomvarphi}), and using the expressions for ${\cal G}^{IJ}$ and $\Gamma^I_{\> JK}$ in the Appendix of \cite{KMSBispectrum}, the equation of motion for the background field $\phi (t)$ takes the form 
\beq
\ddot{\phi} + 3 H \dot{\phi} + \frac{\xi (1 + 6 \xi)}{C} \phi \left( \dot{\phi}^2 + \dot{\chi}^2 \right) - \frac{\dot{f}}{f} \dot{\phi}  + \lambda M_{\rm pl}^4 \frac{\phi ( \phi^2 + \chi^2 )}{2 fC} = 0 .
\label{eomphi}
\eeq
The equation for $\chi$ follows upon replacing $\phi \longleftrightarrow \chi$. Using Eq. (\ref{dotsigma}), the square of the fields' velocity vector becomes
\beq
\dot{\sigma}^2 = \left( \frac{ M_{\rm pl}^2}{2f} \right) \left[ \left( \dot{\phi}^2 + \dot{\chi}^2 \right) + \frac{3 \dot{f}^2}{f} \right] ,
\label{dotsigmaphichi}
\eeq
and the gradient of the potential in the direction $\hat{\sigma}^I$ becomes
\beq
\hat{\sigma}^I V_{, I} = V_{, \sigma} = \frac{\lambda M_{\rm pl}^6  \left( \phi^2 + \chi^2 \right) }{\xi  (2f)^3 } \frac{\dot{f}}{\dot{\sigma}} .
\label{Vsigma2}
\eeq

We may verify that multifield Higgs inflation exhibits slow-roll behavior for typical choices of couplings and initial conditions. First consider the single-field case, in which we set $\chi = \dot{\chi} = 0$. Near the start of inflation (with $\xi \phi^2 \gg M_{\rm pl}^2$), the terms in Eq. (\ref{eomphi}) that stem from the field's noncanonical kinetic term take the form
\beq
\frac{\xi (1 + 6 \xi)}{C} \phi \dot{\phi}^2 - \frac{\dot{f}}{f} \dot{\phi} \rightarrow - \frac{\dot{\phi}^2}{\phi} .
\label{noncanonical}
\eeq
The usual slow-roll requirement for single-field models, $\vert \dot{\phi} \vert \ll \vert H \phi \vert$, ensures that the terms in Eq. (\ref{noncanonical}) remain much less than the $3H \dot{\phi}$ term in Eq. (\ref{eomphi}). Neglecting $\ddot{\phi}$, the single-field, slow-roll limit of Eq. (\ref{eomphi}) becomes
\beq
3H \dot{\phi} \simeq - \frac{\lambda M_{\rm pl}^4}{6 \xi^3 \phi} ,
\eeq
or, upon using $H^2 \simeq V / (3 M_{\rm pl}^2)$,
\beq
\dot{\phi} \simeq - \frac{\sqrt{\lambda} \> M_{\rm pl}^3}{3 \sqrt{3} \> \xi^2 \phi }  .
\label{dotphisinglefield}
\eeq
Setting $\xi = 10^4$ and fixing the initial field velocity by Eq. (\ref{dotphisinglefield}) requires $\phi (0) \geq 0.1 M_{\rm pl}$ to yield $N \geq 70$ efolds of inflation in the single-field case.

A much broader range of initial conditions yields $N \geq 70$ efolds in the two-field case. From Eq. (\ref{Friedmannsigma}) we see that inflation (with $\ddot{a} > 0$) requires $\dot{\sigma}^2 \ll V$. Given the $SO ({\cal N})$ symmetry of the model, we may set $\chi (0) = 0$ without loss of generality, and parameterize the fields' initial velocities as
\beq
\begin{split}
\dot{\phi} (0) &= \frac{\sqrt{\lambda} \> M_{\rm pl}^3}{3 \sqrt{3} \> \xi^2 \phi (0)} \> x , \\
\dot{\chi} (0) &= \frac{\sqrt{\lambda} \> M_{\rm pl}^3}{3 \sqrt{3} \> \xi^2 \phi (0)} \> y 
\end{split}
\label{dotphidotchi}
\eeq
in terms of dimensionless constants $x$ and $y$. (The single-field case corresponds to $x = -1, y = 0$.) Near the start of inflation, when $\xi r^2 = \xi \phi^2 \gg M_{\rm pl}^2$, Eq. (\ref{dotsigmaphichi}) becomes
\beq
\dot{\sigma}^2 \vert_{\chi (0) = 0} \rightarrow \left(\frac{\lambda M_{\rm pl}^4}{4 \xi^2} \right) \left( \frac{M_{\rm pl}^2}{\xi \phi^2 (0) } \right)^2 \frac{4}{27 \xi} \left[ (1 + 6 \xi) x^2 + y^2 \right] .
\label{sigma2field}
\eeq
The first term in parentheses is just the value of the potential, $V$, near the start of inflation, as given in Eq. (\ref{VEflat}). The second term in parentheses is small near the beginning of inflation, given $\xi r^2 \gg M_{\rm pl}^2$. Hence the initial values for $\dot{\phi}$ and $\dot{\chi}$, parameterized by the coefficients $x$ and $y$, may be substantially larger than in the single-field case while still keeping $\dot{\sigma}^2 \ll V$. 

\begin{figure}
\centering
\includegraphics[width=4.5in]{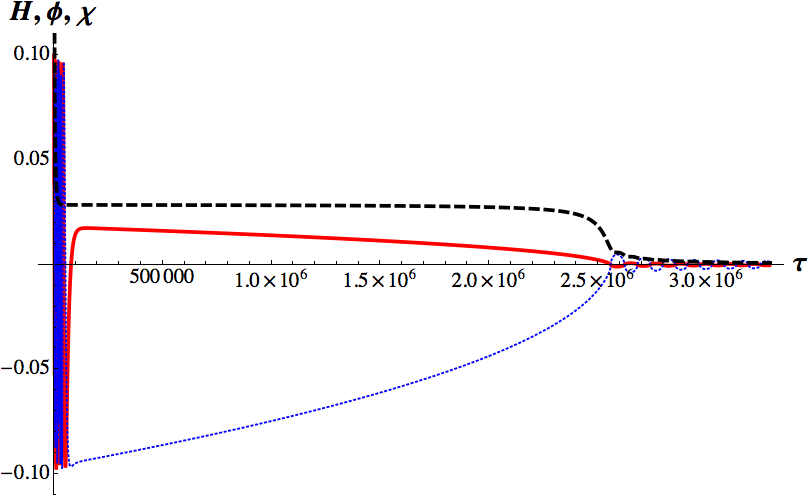}
\caption{\small \baselineskip 14pt The evolution of $H (t)$ (black dashed line) and the fields $\phi (t)$ (red solid line) and $\chi (t)$ (blue dotted line). The fields are measured in units of $M_{\rm pl}$ and we use the dimensionless time variable $\tau = \sqrt{ \lambda} \> M_{\rm pl} t$. We have plotted $10^3 H$ so that its scale is commensurate with the magnitude of the fields. The Hubble parameter begins large, $H (0) = 8.1 \times 10^{-4}$, but quickly falls by a factor of 30 as it settles to its slow-roll value of $H = 2.8 \times 10^{-5}$. Inflation proceeds for $\Delta \tau = 2.5 \times 10^6$ to yield $N = 70.7$ efolds of inflation. The solutions shown here correspond to $\xi = 10^4$, $\phi (0) = 0.1$, $\chi (0) = 0$, $\dot{\phi} (0) = - 2 \times 10^{-6}$, and $\dot{\chi} (0) = 2 \times 10^{-2}$. For the same value of $\phi (0)$, Eq. (\ref{dotphisinglefield}) corresponds to $\dot{\phi} (0) = - 2 \times 10^{-8}$ for the single-field case. }
\label{HfieldsHiggs}
\end{figure}

Fig. \ref{HfieldsHiggs} shows $H (t)$, $\phi (t)$, and $\chi (t)$ for a scenario in which $\dot{\phi} (0)$ and $\dot{\chi} (0)$ greatly exceed the single-field relation of Eq. (\ref{dotphisinglefield}): $\vert x \vert = 10^2$ and $\vert y \vert = 10^6$. As is evident in the figure, the large initial velocities cause the fields to oscillate rapidly. The extra kinetic energy makes the initial value of $H (t)$ larger than in the corresponding single-field case. The increase in $H$, in turn, causes the fields' velocities to damp out even more quickly, due to the $3H \dot{\phi}$ and $3 H \dot{\chi}$ Hubble-drag terms in each field's equation of motion. Thus the system rapidly settles into a slow-roll regime that continues for 70 efolds. As shown in Fig. 3, we may achieve $N \geq 70$ efolds with even smaller initial field values by making the initial field velocities correspondingly larger. 

\begin{figure}
\centering
\includegraphics[width=3in]{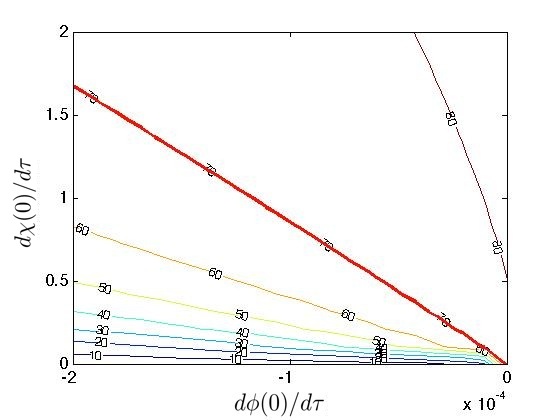} \includegraphics[width=3in]{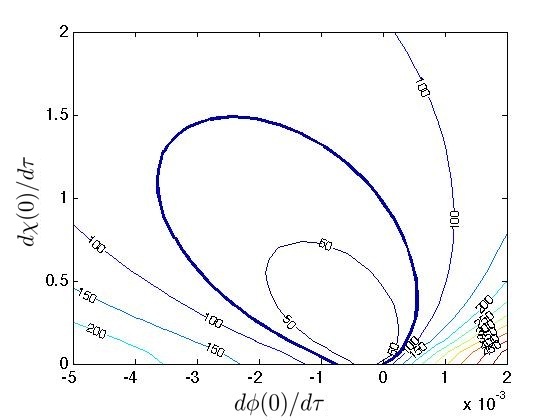}\\
\includegraphics[width=3in]{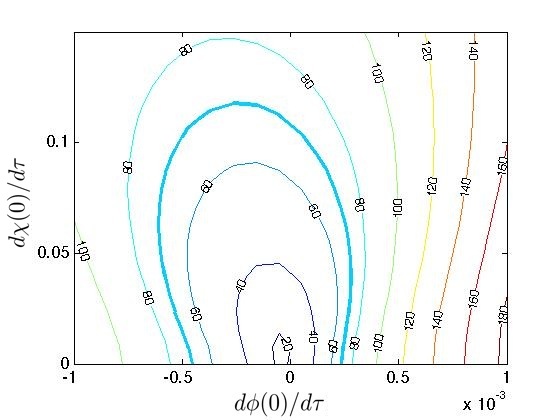} \includegraphics[width=3in]{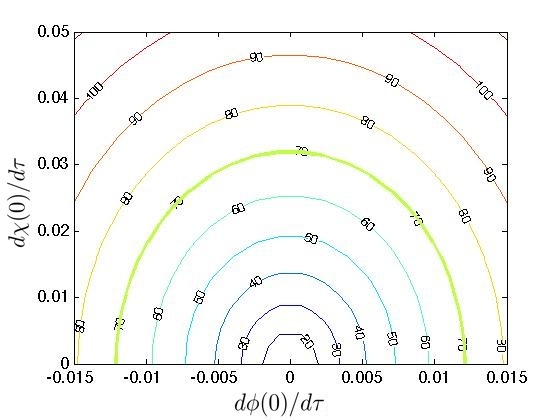}
\caption{\small \baselineskip 14pt Contour plots showing the number of efolds of inflation as one varies the fields' initial conditions, keeping $\xi = 10^4$ fixed. In each panel, the vertical axis is $\dot{\chi} (0)$ and the horizontal axis is $\dot{\phi} (0)$. The panels correspond to $\phi (0) = 10^{-1} \>M_{\rm pl}$ (top left), $10^{-2} \> M_{\rm pl}$ (top right), $5 \times 10^{-3} \> M_{\rm pl}$ (bottom left) and $10^{-4} \> M_{\rm pl}$ (bottom right), and we again use dimensionless time $\tau = \sqrt{\lambda} \> M_{\rm pl} t$. In each panel, the line for $N = 70$ efolds is shown in bold. Note how large these initial velocities are compared to the single-field expectation of Eq. (\ref{dotphisinglefield}).}
\label{fig3}
\end{figure}

\section{Turn Rate}

\begin{figure}
\centering
\includegraphics[width=3in]{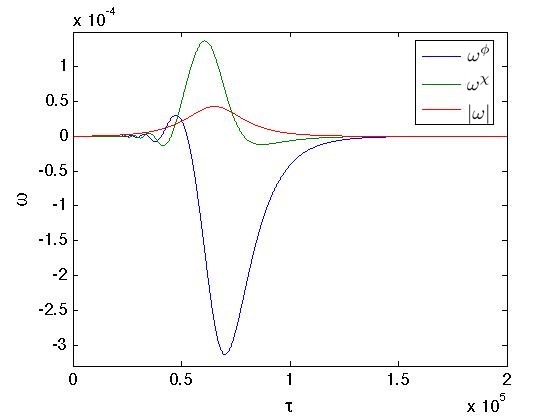} \includegraphics[width=3in]{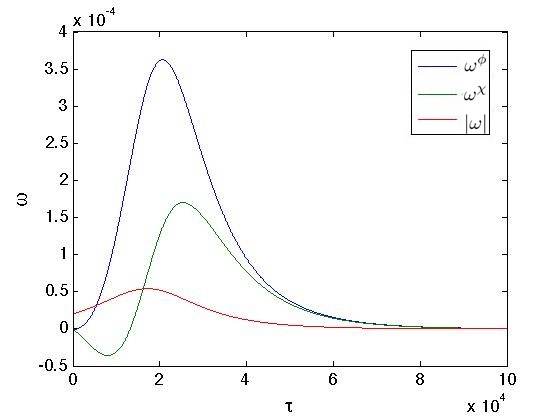}
\caption{\small  \baselineskip 14pt Evolution of the turn rate. The left picture shows the evolution with initial conditions as in Fig. \ref{HfieldsHiggs}. The right figure has initial conditions $\phi(0)=0.1$, $\chi(0)=\dot \phi(0) = 0$, and $\dot \chi (0) = 2 \times 10^{-5}$ in units of $M_{\rm pl}$ and $\tau = \sqrt{\lambda} \> M_{\rm pl} t$. In both cases we set $\xi = 10^4$. Recall from Fig. \ref{HfieldsHiggs} that inflation lasts until $\tau_{\rm end} \sim {\cal O} ( 10^6 )$ for these initial conditions; hence we find that $\omega$ damps out within a few efolds after the start of inflation.}
\label{omegaplot}
\end{figure}

The components of the turn-rate, $\omega^I$ in Eq. (\ref{omegadef}), take the form
\beq
\omega^\phi = - \frac{\lambda M_{\rm pl}^4}{\dot{\sigma}} \frac{ r^2 }{2f} \left[ \frac{\phi}{C} - \frac{M_{\rm pl}^2}{4 f^2} \frac{\dot{\phi}}{\dot{\sigma}^2} \left( \phi \dot{\phi} + \chi \dot{\chi} \right) \right] .
\label{alphaphi}
\eeq
The other component, $\omega^\chi$, follows upon replacing $\phi \longleftrightarrow \chi$. The length of the turn-rate vector is given by
\beq
\omega = \vert \omega^I \vert = \sqrt{ {\cal G}_{IJ} \omega^I \omega^J } = \frac{1}{\dot{\sigma}} \sqrt{ \hat{s}^{KM} V_{, K} V_{, M} } \> ,
\label{alphalength1}
\eeq
where the final expression follows upon using the definition of $\omega^I$ in Eq. (\ref{omegadef}) and the identity $\hat{s}^{KM} = \hat{s}^K_{\>\> A} \hat{s}^{MA}$, which follows from Eq. (\ref{sIJ}). We find
\beq
\dot{\sigma}^2 \omega^2 = \hat{s}^{KM} V_{, K} V_{, M} = \frac{\lambda^2 M_{\rm pl}^{10}}{(2f)^5 C} r^6 \left[ C - \xi^2 r^2 \right] - \left( V_{, \sigma} \right)^2 .
\label{alphalength2}
\eeq
The evolution of the turn rate for typical initial conditions is shown in Fig. \ref{omegaplot}.

In order to analyze the evolution of the background fields, it is easier to move from Cartesian to polar coordinates, in which the angular velocity and turn rate have more intuitive behavior. In addition to the radius, $r^2 = \phi^2+\chi^2$, we also define the angle 
\beq
\gamma \equiv {\rm arctan} \left ( {\chi \over \phi} \right ). 
\label{gamma}
\eeq
Single-field trajectories correspond to constant $\gamma(t)$. In the polar coordinate system, the background dynamics of Eq. (\ref{Friedmannsigma}) may be written
\beq
\begin{split}
H^2 &= \frac{1}{12f} \left[  \dot{r}^2 +  r^2\dot  \gamma^2+ {3\xi^2 \over f} r^2 \dot r^2 + {\lambda M_{\rm pl}^2 \over 2} {r ^4 \over ( M_{\rm pl}^2 + \xi r^2 )}  \right] , \\
\dot{H} &= - \frac{1}{4f} \left[  \dot{r}^2 +  r^2 \dot \gamma^2 + {3\xi^2 \over f} r^2 \dot r^2\right] .
\end{split}
\label{polar_H}
\eeq
The equations of motion become
\beq
\ddot{r} + 3 H \dot{r} - r \dot \gamma^2 + {\xi (1+6\xi) \over C} r \left ( \dot r^2 + r^2 \dot \gamma^2 \right ) -{\xi \over f} \dot r^2 r + \lambda M_{\rm pl}^4 {r^3 \over 2fC}  = 0 
\label{eomr}
\eeq
and
\beq
\ddot{\gamma} + \left ( 3 H  + 2 {\dot r \over r}   {M_{\rm pl}^2 \over (M_{\rm pl}^2+\xi r^2 ) }\right )\dot \gamma  = 0 .
\label{polar_eom}
\eeq

In this new basis the turn rate may be written compactly as
\beq
\omega^2 ={\lambda^2 M_{\rm pl}^8 \over 2fC} \left ( \frac{r^4 \dot \gamma }{r^2 \dot \gamma^2 (M_{\rm pl}^2 +\xi r^2) + \dot r ^2 C }\right )^2 
\eeq
This expression vanishes in both the limits $\vert \dot{\gamma} \vert \rightarrow 0$ and $\vert \dot{\gamma} \vert \rightarrow \infty$: if the angular velocity is either too large or too small, the fields' evolution reverts to effectively single-field behavior (either purely radial motion or purely angular motion). Of the two limits, however, only pure-radial motion is stable. It is ultimately the evolution of $\gamma(t)$ that will determine the fate of the turn rate.

It is obvious from Eq. (\ref{polar_eom}) that the line $\dot \gamma =0$ is the fixed point of the angular motion. The character of the fixed point is defined by the sign of the $\dot \gamma$ term, which is less trivial. It can be negative close to $r=0$ due to the high curvature of the field manifold and the small value of the Hubble parameter, but in the slow-roll regime of the radial field, with $\xi r^2 \gg M_{\rm pl}^2$, the sign of $\dot{\gamma}$ is safely positive. That means that we can treat the angular motion as damped throughout inflation.

For large nonminimal coupling and/or slow rolling of the radial field the last term in Eq. (\ref{polar_eom}) may be neglected, which yields
\beq
\ddot{\gamma} +  3 H\dot \gamma = 0.
\label{ddotgamma1}
\eeq
The only complicated object in Eq. (\ref{ddotgamma1}) is the Hubble parameter, which may be simplified in the limit of a slow rolling radial field and large nonminimal coupling upon making use of Eq. (\ref{polar_H}):
\beq
H \simeq {1\over \sqrt{6 \xi}} \sqrt{\dot \gamma^2  +\frac{\lambda M_{\rm pl}^2}{ 2\xi}} .
\label{Happrox}
\eeq
Then Eq. (\ref{ddotgamma1}) becomes 
\beq
\ddot{\gamma} +  {3\over \sqrt{6 \xi}} \sqrt{\dot \gamma^2  + \frac{ \lambda M_{\rm pl}^2}{ 2\xi} } \> \dot \gamma \simeq 0.
\label{ddotgammaSR}
\eeq
Although Eq. (\ref{ddotgammaSR}) can be solved exactly (see the Appendix), it is instructive to examine the two limits of large and small $\dot{\gamma}$, which provide most of the relevant information. 

For small angular velocity, $\dot \gamma \ll \sqrt{ \lambda M_{\rm pl}^2 /  2\xi}$, we recover the linear limit
\beq
\ddot \gamma + \frac{3}{\xi} \sqrt{\frac{\lambda M_{\rm pl}^2}{12} } \>  \dot \gamma \simeq 0
\eeq
with the solution 
\beq
\dot \gamma = \dot \gamma _0 \> \exp \left[ - \frac{ \sqrt{3\lambda} }{2 \xi} M_{\rm pl} \> t \right] \propto e^{-3N} ,
\label{dotgammalinear}
\eeq
where $N=Ht$. It is very easy to measure time in efolds in this limit, since the Hubble parameter is nearly constant. Eq. (\ref{dotgammalinear}) illustrates that any small, initial angular velocity will be suppressed within a couple of efolds, or equivalently within a time of the order of $\xi /  ( \sqrt{\lambda} \>M_{\rm pl} )$.

In the opposite limit, $\dot \gamma \gg  \sqrt{\lambda M_{\rm pl}^2 /  2\xi}$, which we call the nonlinear regime, Eq. (\ref{ddotgammaSR}) becomes
\beq
\ddot \gamma + 3 {1\over \sqrt{6 \xi }}  \dot \gamma^2 \simeq 0
\eeq
with the solution 
\beq 
\dot \gamma = \left[ \frac{1}{ \dot{\gamma}_0} + \frac{3t}{\sqrt{6 \xi} } \right]^{-1} .
\label{dotgammanonlinear}
\eeq
Given Eqs. (\ref{dotgammalinear}) and (\ref{dotgammanonlinear}), we may follow the evolution of any initial angular velocity. If $\dot{\gamma}$ begins large enough it will start in the nonlinear regime, where it will stay until it becomes of order  $ \sqrt{ \lambda M_{\rm pl}^2 /  2\xi} $. We parameterize the cross-over regime as
\beq
\dot{\gamma} = \sqrt{\lambda} M_{\rm pl} \> \frac{z}{\sqrt{2\xi} } 
\label{gammacrossover}
\eeq
where $z$ is some constant of order one. The cross-over time may then be estimated by inverting Eq. (\ref{dotgammanonlinear}) to find
\beq
t_{\rm nl}={\sqrt{6\xi} \over 3} \left[ \sqrt{ \frac{ 2 \xi }{\lambda} } \frac{1}{M_{\rm pl} \> z } - {1\over \dot \gamma_0} \right] .
\label{tnl}
\eeq
There exists an upper limit on the time it takes for the angular velocity to decay, namely
\beq 
\sqrt{\lambda} M_{\rm pl} \> t_{\rm nl,max}={2 \over \sqrt{3}}  { \xi \over z} .
\label{tnlmax}
\eeq

\begin{figure}
\centering
\includegraphics[width=3in]{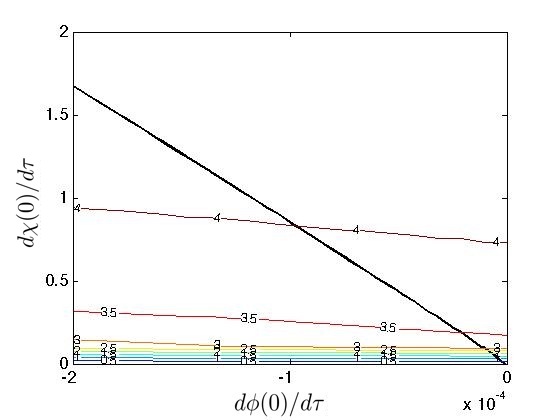} \includegraphics[width=3in]{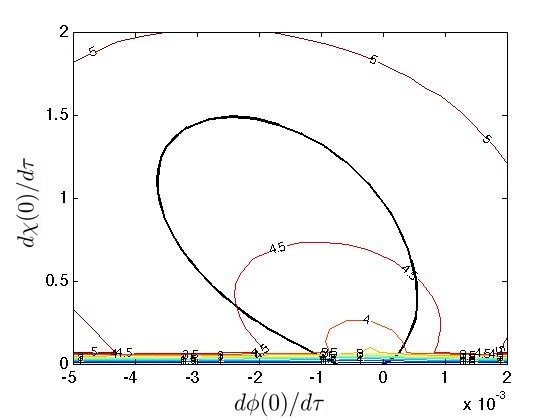}\\
\includegraphics[width=3in]{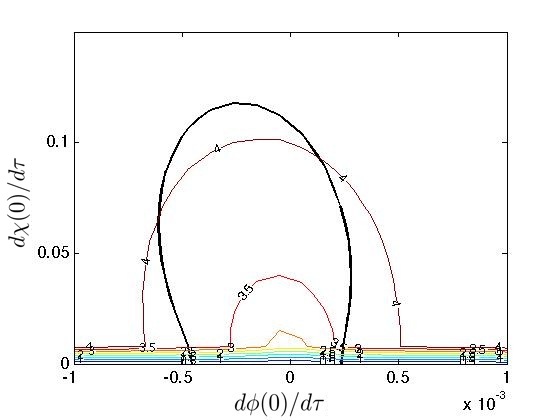} \includegraphics[width=3in]{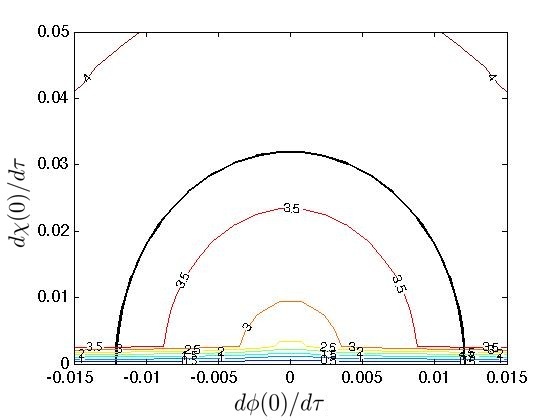}
\caption{\small \baselineskip 14pt Contour plots showing the number of efolds at which the maximum of the turn rate occurs, as one varies the fields' initial conditions. In each panel, the vertical axis is $\dot{\chi} (0)$ and the horizontal axis is $\dot{\phi} (0)$. The panels correspond to $\phi (0) = 10^{-1} \>M_{\rm pl}$ (top left), $10^{-2} \> M_{\rm pl}$ (top right), $5 \times 10^{-3} \> M_{\rm pl}$ (bottom left) and $10^{-4} \> M_{\rm pl}$ (bottom right). We set $\xi = 10^4$ and use the dimensionless time-variable $\tau = \sqrt{\lambda} \> M_{\rm pl} t$. The thick black curve is the contour line of initial conditions that yield $N = 70$ efolds.}
\label{omegacontour}
\end{figure}

\begin{figure}
\centering
\includegraphics[width=3in]{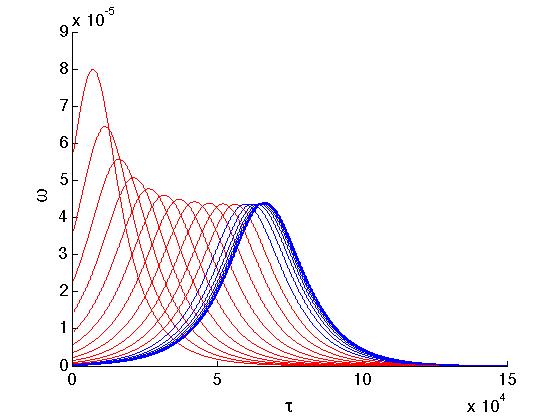} 
\caption{\small \baselineskip 14 pt The turn rate as a function of time for different values of the initial angular velocity. The parameters used are $\xi = 10^4$, $\phi(0)=0.1 M_{\rm pl}$, $\dot \phi(0)=\chi(0)=0$, and $ {0.01\over \sqrt{2\xi}} \leq \dot \gamma(0) \leq {100 \over \sqrt{2\xi}}$, in terms of dimensionless time, $\tau = \sqrt{\lambda} \> M_{\rm pl} t$. In these units and for $\phi (0) = 0.1 M_{\rm pl}$, inflation lasts until $\tau_{\rm end} \sim {\cal O} (10^6)$.   }
\label{omega_sweep}
\end{figure}

We have verified all of these analytic predictions using numerical calculations of the exact equations for the coupled two-field system in an expanding universe. In Fig. \ref{omegacontour} we plot the number of efolds from the beginning of inflation at which the turn rate reaches its maximum value, as we vary the fields' initial velocities. Note that for any combination of initial conditions that yields at least $N_{\rm tot} = 70$ efolds, $\omega$ reaches its maximum value between $N (\omega_{\rm max}) = 3.5$ and $5$ efolds from the start of the fields' evolution (for the range of initial conditions considered there). In Fig. \ref{omega_sweep} we plot $\omega$ as a function of time as we vary the initial angular velocity, $\dot{\gamma} (0)$. The curves in red correspond to initial conditions in the linear regime, while the curves in blue start in the nonlinear regime. Note that the curves starting in the nonlinear regime have the same amplitude. The existence of a maximum time, $t_{\rm nl, max}$, is evident from the bunching of the blue curves. We find $\sqrt{\lambda} M_{\rm pl}\> t_{\rm nl, max} = \tau_{\rm nl, max} \sim {\rm few} \times \xi \sim 10^4$, as expected from Eq. (\ref{tnlmax}). In these units and for the initial conditions used in Fig. \ref{omega_sweep}, inflation lasts until $\tau_{\rm end} \sim {\cal O} (10^6)$, so $\tau_{\rm nl, max}$ occurs very early after the onset of inflation.

\begin{figure}
\centering
\includegraphics[width=3in]{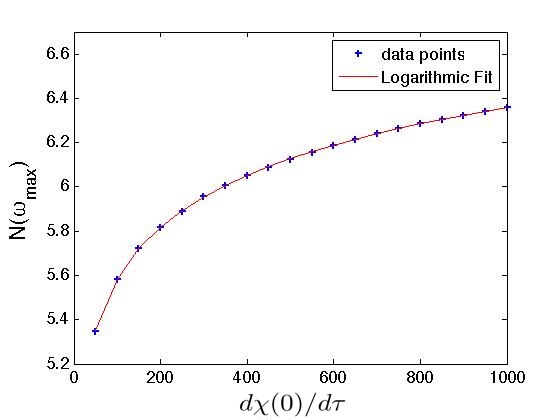} \includegraphics[width=3in]{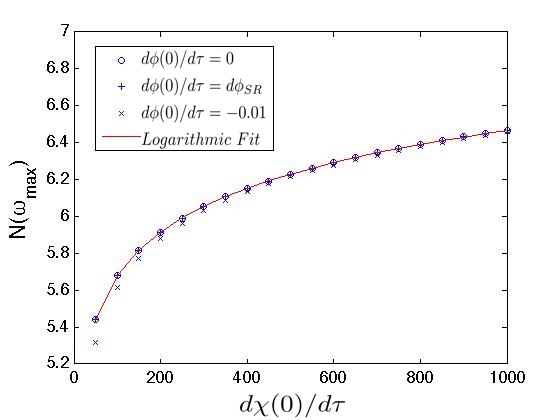}
\caption{\small \baselineskip 14pt Number of efolds until the maximum value of the turn rate is reached, as a function of $\dot{\chi} (0)$. On the left we plot $N (\omega_{\rm max})$ for initial conditions that yield $N_{\rm tot} = 70$ efolds; on the right we plot the same quantity for various values of $\dot{\phi} (0)$. The logarithmic fit is an excellent match to our analytic result, Eq. (\ref{Nnl}).}
\label{omegaN}
\end{figure}

Eq. (\ref{dotgammalinear}) shows that the linear region lasts at most a few efolds, so the duration of the nonlinear region is what will ultimately determine whether or not multifield effects will persist until observationally relevant length scales first cross the Hubble radius. In the nonlinear regime, Eq. (\ref{Happrox}) yields $H \simeq \dot{\gamma} / \sqrt{6 \xi}$ with $\dot{\gamma}$ given by Eq. (\ref{dotgammanonlinear}). The number of efolds for which the nonlinear regime persists is given by
\beq
N_{\rm nl} = \int _0 ^{t_{\rm nl}} H dt \simeq  {1\over \sqrt{6\xi}}  \int _0 ^{t_{\rm nl}}  \dot \gamma dt = {1 \over 3 } \ln \left ( \sqrt{\frac{2\xi}{\lambda}} \frac{\dot{\gamma}_0}{ M_{\rm pl} \> z} \right ) .
\label{Nnl}
\eeq
We examine Eq. (\ref{Nnl}) numerically by fixing $\xi = 10^4$ and $\phi (0) = 0.1 M_{\rm pl}$ and choosing pairs of initial velocities, $\dot{\phi} (0)$ and $\dot{\chi} (0)$, that yield 70 efolds (see Fig. \ref{omegaN}, left); and also by setting $\dot{\phi} (0)$ to various constant values and varying $\dot{\chi} (0)$ (Fig. \ref{omegaN}, right). The results fall neatly along a least-squares logarithmic fit, as expected from Eq. (\ref{Nnl}). The function $N_{\rm nl}$ grows slowly. In order for multifield effects to remain important more than a few efolds after the start of inflation, the initial angular velocity would need to be enormous: at least ten orders of magnitude larger than typical values of the initial field velocity for single-field inflation, as given in Eq. (\ref{dotphisinglefield}). We do not know of any realistic mechanism that could generate initial field velocities so large. Moreover, for many combinations of initial conditions shown in the righthand side of Fig. \ref{omegaN}, $N_{\rm tot} > 70$ efolds (several sets of initial conditions yield $N_{\rm tot} \sim 90$ efolds). For those scenarios, the turn rate reaches its maximum value deep within the early phase of the system's evolution, long before observationally relevant perturbations first cross the Hubble radius. The multifield dynamics for this model thus behave similarly to those in related multifield models of inflation that involve the Higgs sector, such as \cite{JGB}.

We may consider the behavior of $a(t)$ and $H(t)$ in the two different regimes more closely. From the definition of $H$ and $\dot H$ in Eq. (\ref{polar_H}) and neglecting the terms proportional to $\dot r$ (which is equivalent to requiring the field to be slow rolling along the radial direction), we find 
\beq
 {\ddot a \over a}  = \dot H +H^2= {1\over 12 f} \left ( -2 r^2\dot \gamma ^2 + {\lambda M_{\rm pl}^2 \over 2} {r^4 \over M_{\rm pl}^2 + \xi r^2} \right ) .
\eeq
When the potential dominates we recover what we called the linear regime in the analysis of the decay of $\omega$. In that regime
\beq
 {\ddot a \over a} > 0 ,
 \eeq
which is an accelerated expansion or cosmological inflation. However, in the nonlinear regime, when $\dot \gamma$ dominates, the situation reverses and we find
\beq
 {\ddot a \over a} = -{1\over 6f} r^2 \dot \gamma^2 < 0 ,
 \eeq
which is an expansion and a very rapid one (because of the large value of $H$), but it is not inflation. Regardless of whether we have true inflation or simply rapid expansion at early times, we may always define the number of efolds as
\beq
N = \int_{t_{\rm in}} ^{t_{\rm end}} H dt .
\eeq
Thus we may use $N$ as our clock and measure time in efolds from the beginning of the system's evolution, regardless of whether it is in the inflationary phase or not. The fact that in the nonlinear regime the universe is not inflating only makes our results stronger:  all multifield effects decay before the observable scales exit the horizon in a model that produces enough inflation to solve the standard cosmological problems.

\begin{figure}
\centering
\includegraphics[width=3in]{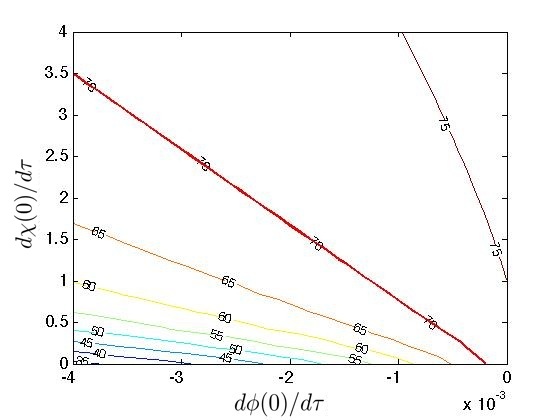} \includegraphics[width=3in]{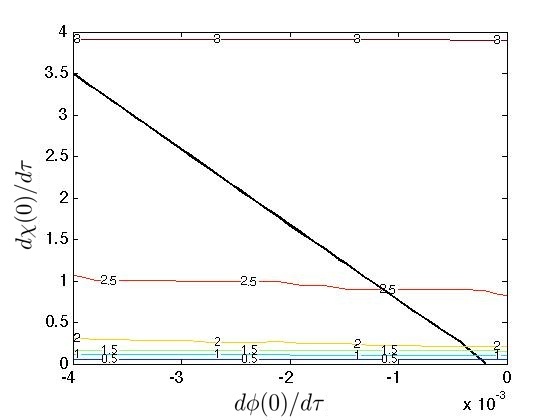}\\
\includegraphics[width=3in]{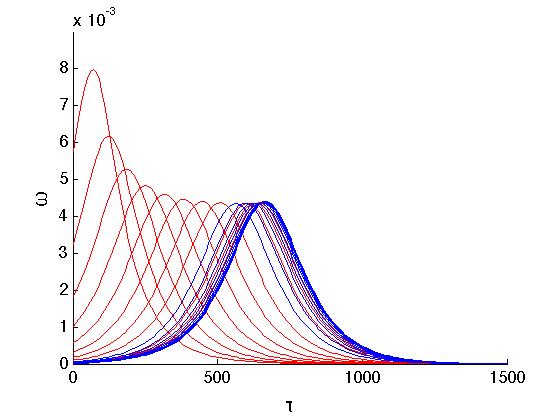} \includegraphics[width=3in]{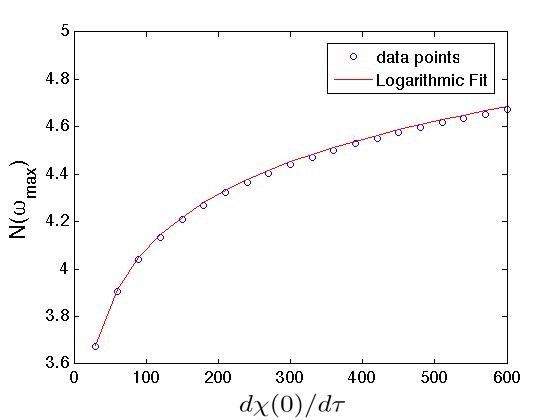}
\caption{\small \baselineskip 14pt Dynamics of our two-field model with $\xi=10^2$, $\phi(0)=1 M_{\rm pl}$, and $\chi(0)=0$. Clockwise from top left: 
 (1) Contour plot showing the number of efolds as one varies the fields' initial conditions. The thick curve corresponds to $70$ efolds. (2) Contour plot showing the number of efolds at which the maximum of the turn rate occurs, as one varies the fields' initial conditions. The thick curve corresponds to $N_{\rm tot} = 70$ efolds. (3) Number of efolds until the maximum value of the turn rate is reached for initial conditions giving $N_{\rm tot} = 70$ efolds, along with a logarithmic fit. (4) The turn rate as a function of time for different values of the initial angular velocity, with $\dot \phi(0)=0$ and $ {0.01\over \sqrt{2\xi}} \leq \dot \gamma(0) \leq {100 \over \sqrt{2\xi}}$, in units of $\tau = \sqrt{\lambda} \> M_{\rm pl} t$.
}
\label{fig:ksi100}
\end{figure}

As a final test of our analysis we set $\xi=10^2$ instead of $\xi = 10^4$. The smaller value of the nonminimal coupling does not lead to a viable model of Higgs inflation --- the WMAP normalization of the power spectrum requires a larger value of $\xi$ \cite{Higgsinfl} --- but we may nonetheless study the dynamics of such a model. We collect the important information about the dynamics of this model in Fig. \ref{fig:ksi100}. As expected, the model can provide $70$ or more efolds of inflation for a wide range of parameters, and the corresponding turn rate peaks well before observationally relevant length scales first crossed the Hubble radius, even when we increase $\dot{\chi} (0)$ to a few hundred in units of $\tau = \sqrt{\lambda} M_{\rm pl} t$. The excellent logarithmic fit of the time at which the turn rate is maximum versus $\dot{\chi} (0)$ is again evident. Finally the curves of the turn rate versus time show the same qualitative and quantitative characteristics as Fig. \ref{omega_sweep} for $\xi=10^4$. Specifically, if one rescales time and the turn rate appropriately by $\xi$, the two sets of curves would be hardly distinguishable. 

\section{Implications for the Primordial Spectrum}

We have found that in models with an $SO ({\cal N})$ symmetry among the scalar fields, the turn rate quickly damps to negligible magnitude within a few efolds after the start of inflation. In this section we confirm that such behavior yields empirical predictions for observable quantities like the primordial power spectrum of perturbations that reproduce expectations from corresponding single-field models. 

For models that behave effectively as two-field models, which includes the class of $SO ({\cal N})$-symmetric models we investigate here, we may distinguish two scalar perturbations: the perturbations in the adiabatic direction, $Q_\sigma$ defined in Eq. (\ref{Qsigmadeltas}), and a scalar entropic perturbation \cite{KMSBispectrum},
\beq
Q_s \equiv \frac{\omega_I}{\omega} \delta s^I .
\label{Qs}
\eeq
We noted in Eq. (\ref{RcQsigma}) that $Q_\sigma$ is proportional to the gauge-invariant curvature perturbation, ${\cal R}_c$. We adopt a similar normalization for the entropy perturbation,
\beq
{\cal S} \equiv \frac{H}{\dot{\sigma}} Q_s .
\label{S}
\eeq
In the long-wavelength limit, the adiabatic and entropic perturbations obey \cite{transferfunction,KMSBispectrum}
\beq
\begin{split}
\dot{\cal R}_c &= \alpha H {\cal S} + {\cal O} \left( \frac{k^2}{a^2 H^2} \right) , \\
\dot{\cal S} &= \beta H {\cal S} + {\cal O} \left( \frac{k^2}{a^2 H^2} \right) ,
\end{split}
\label{dotRdotS}
\eeq
so that we may define the transfer functions
\beq
\begin{split}
T_{\cal RS} (t_*, t) &= \int_{t_*}^t dt' \>  \alpha (t') H (t') T_{\cal SS} (t_*, t') , \\
T_{\cal SS} (t_*, t) &= \exp \left[ \int_{t_*}^t dt' \> \beta (t') H (t') \right] .
\end{split}
\label{TrsTss}
\eeq
We take $t_*$ to be the time when a fiducial scale of interest first crossed the Hubble radius during inflation, defined by $a^2 (t_*) H^2 (t_*) = k_*^2$. In \cite{KMSBispectrum}, we calculated
\beq
\alpha (t) = \frac{2\omega (t)}{H (t)}
\label{alpha}
\eeq
and
\beq
\beta (t) = - 2 \epsilon - \eta_{ss} + \eta_{\sigma\sigma} - \frac{4}{3} \frac{\omega^2}{H^2} ,
\label{beta}
\eeq
where $\epsilon \equiv - \dot{H} / H^2$ and the other slow-roll parameters are defined as
\beq
\begin{split}
\eta_{\sigma\sigma} &\equiv M_{\rm pl}^2 \frac{ {\cal M}_{\sigma\sigma}}{V} , \\
\eta_{ss} &\equiv M_{\rm pl}^2 \frac{ \omega_I \omega^J {\cal M}^I_{\>\> J} }{ \omega^2 V} .
\end{split}
\label{etas}
\eeq
The dimensionless power spectrum is given by
\beq
{\cal P}_{\cal R} = \frac{k^3}{2 \pi^2} \vert {\cal R}_c \vert^2
\eeq
and hence, from Eqs. (\ref{dotRdotS}) and (\ref{TrsTss}),
\beq
{\cal P}_{\cal R} (k) = {\cal P}_{\cal R} (k_*) \left[ 1 + T_{\cal RS}^2 (t_*, t) \right] ,
\label{PR}
\eeq
where $k$ corresponds to a length scale that crossed the Hubble radius at some time $t > t_*$. The spectral index is then given by
\beq
n_s (t) = n_s (t_*) - \left[ \alpha (t) + \beta(t) T_{\cal RS} (t, t_*) \right] \sin (2\Delta) ,
\label{ns}
\eeq
where 
\beq
\cos \Delta \equiv \frac{ T_{\cal RS}}{\sqrt{ 1 + T_{\cal RS}^2 }} .
\label{Delta}
\eeq
In the limit $( \omega  / H ) \ll \eta_{\sigma\sigma}$, the spectral index evaluated at $N_*$ assumes the single-field form \cite{BTW,MalikWands,DKnperts},
\beq
n_s (t_*) = 1 - 6 \epsilon (t_*) + 2 \eta_{\sigma\sigma} (t_*) .
\label{nsstar}
\eeq

Crucial to note is that the turn rate, $\omega$, serves as a window function within $T_{\cal RS} (t, t_*)$: once the coefficient $\alpha = 2 \omega / H$ becomes negligible, there will effectively be no transfer of power from the entropic to the adiabatic perturbations, much as we had found by examining the source terms on the righthand sides of Eqs. (\ref{eomQsigma}) and (\ref{eomdeltas}). The question then becomes whether $\omega (t)$, and hence $T_{\cal RS} (t_*, t)$, can depart appreciably from zero at times when perturbations on length scales of observational interest first cross the Hubble radius.

The longest length scales of interest are often taken to be those that first crossed the Hubble radius $N_* = 55 \pm 5$ efolds before the end of inflation \cite{MFB,BTW,MalikWands}. Closer analysis suggests that length scales that first crossed the Hubble radius $N_* = 62 - 63$ efolds before the end of inflation correspond to the size of the present horizon \cite{Ncross}. Meanwhile, we follow \cite{BTW} in assuming that successful inflation requires $N_{\rm tot} \geq 70$ efolds to solve the horizon and flatness problems. The question then becomes whether $\omega (t)$, and hence $T_{\cal RS} (t_*, t)$, can differ appreciably from zero for $N_* \leq 63$. Given the analysis in Section IV, the best chance for this to occur is for initial conditions that produce the minimum amount of inflation, $N_{\rm tot} = 70$. 

In Table I, we present numerical results for key measures of multifield dynamics. In each case we set $\xi = 10^4$, $\phi (0) = 0.1 M_{\rm pl}$, and $\chi (0) = 0$. We vary $\dot{\chi  } (0)$ as shown and adjust $\dot{\phi} (0)$ in each case so as to produce exactly $N_{\rm tot} = 70$ efolds of inflation. Because $T_{\cal RS}$ remains so small in each of these cases, there is no discernible running of the spectral index within the window $N_* = 63$ to $N_* = 40$ efolds before the end of inflation. If we consider a fiducial scale $k_*$ that first crosses the Hubble radius at $N_* = 63$ efolds before the end of inflation, then we find $n_s = 0.97$ across the whole range of initial conditions, in excellent agreement with the measured value of $n_s = 0.971 \pm 0.010$ \cite{WMAP9}. If instead we set $k_*$ as the scale that first crossed the Hubble radius $N_* = 60$ efolds before the end of inflation, we find $n_s = 0.967$ across the entire range of initial conditions, again in excellent agreement with the latest measurements.

\begin{table}[t]
\centering
\begin{tabular}{|c|c|c|c|c|}
\hline
$\dot{\chi} (0)$ & $\omega (N_* = 63)$ & $T_{\cal RS} (\rm max)$ & $n_s (N_* = 63)$ & $n_s (N_* = 60)$ \\
\hline
$10^{-2}$ & $\>1.16 \times 10^{-10} \>$ & $\>2.68 \times 10^{-6} \>$ & $0.969$ & $0.967$ \\
$10^{-1}$ & $\>1.20 \times 10^{-9}\>$ & $\>2.76 \times 10^{-5}\>$ & $0.969$ & $0.967$ \\
$1$ & $\>9.41 \times 10^{-9} \>$ & $\>2.18 \times 10^{-4}\>$ & $0.969$ & $0.967$ \\
$10^1$ & $\>1.18 \times 10^{-7} \>$ & $\>2.72 \times 10^{-3} \>$ & $0.969$ & $0.967$ \\
$10^2$ & $\>1.12 \times 10^{-6} \>$ & $\>2.59 \times 10^{-2} \>$ & $0.973$ & $0.967$ \\
\hline
\end{tabular}
\caption{\small \baselineskip 14pt Numerical results for measures of multifield dynamics for Higgs inflation with $\xi = 10^4$. We use dimensionless time $\tau = \sqrt{\lambda} M_{\rm pl} t$. } 
\end{table}

\section{Conclusions}

In this paper we have analyzed Higgs inflation as a multifield model with nonminimal couplings. Because the Goldstone modes of the Standard Model electroweak Higgs sector remain in the spectrum at high energies in renormalizable gauges, we have incorporated their effects in the dynamics of the model. Because of the high symmetry of the Higgs sector --- guaranteed by the $SU (2)$ electroweak gauge symmetry, which manifests as an $SO (4)$ symmetry among the scalar fields of the Higgs sector --- the nonmiminal couplings for the various scalar fields take precisely the same value ($\xi_\phi = \xi_\chi = \xi$), as do the tree-level couplings in the Jordan-frame potential ($\lambda_\phi = \lambda_\chi = \lambda$, and so on). The effective potential in the Einstein frame therefore contains none of the irregular features, such as bumps or ridges, that were highlighted in \cite{KMSBispectrum} for the case of multiple fields with arbitrary couplings. With no features such as ridges off of which the fields may fall during their evolution, Hubble drag will always cause any initial angular motion within field space to damp out rapidly. Increasing the initial angular velocity to arbitrarily large values --- well into what we call the nonlinear regime --- only increases the value of $H$ at early times, which makes the Hubble drag even more effective and hence hastens the damping out of the multifield effects. 

The rapidity with which the turn-rate damps to zero combined with the requirement of $N_{\rm tot} \geq 70$ efolds for successful inflation means that the multifield dynamics become negligible before perturbations on scales of observational relevance first cross the Hubble radius. Even if we push the observational window of interest back to $N_* = 63$ efolds before the end of inflation, rather than the usual assumption of $N_* = 55 \pm 5$, we find that the model relaxes to effectively single-field dynamics prior to $N_*$. Hence the predictions from Higgs inflation for observable quantities, such as the spectral index of the power spectrum of primordial perturbations, reduce to their usual single-field form. Moreover, the absence of multifield effects for times later than $N_*$ means that this model should produce negligible non-Gaussianities during inflation, in contrast to the broader family of models studied in \cite{KMSBispectrum}.

The methods we introduce here may be applied to any multifield model with nonminimal couplings and an $SO ({\cal N})$ symmetry among the fields in field space. The conclusion therefore appears robust that such highly symmetric models should behave effectively as single-field models, at least within the observational window of interest between $N_* = 63$ and $N_* = 40$ efolds before the end of inflation. Of course, multfield effects could become important in such models at the end of inflation, during epochs such as preheating \cite{Higgspreheat}. Such processes remain under study.

\section*{ Appendix A: Angular evolution of the field}

For completeness, let us integrate the angular equation of motion, Eq. (\ref{ddotgammaSR}), for all values of $\dot \gamma$ (in the slow roll regime of the radial field). This yields
\beq
\frac{\dot\gamma(t) \left( \sqrt{ \lambda }  M_{\text{pl}} +   \sqrt{2 \xi \dot\gamma_0^2+\lambda  M_{\text{pl}}^2 }\right)}{\dot\gamma_0 \left( \sqrt{ \lambda}   M_{\text{pl}} + \sqrt{2 \xi \dot\gamma^2 (t) + \lambda  M_{\text{pl}}^2 }\right)} = \exp \left[ - \frac{\sqrt{3 \lambda} \> M_{\rm pl} t}{2 \xi} \right] .
\label{gammasol}
\eeq
In the two limits,  $  \dot \gamma_0 \ll \sqrt{\lambda} M_{\rm pl} / \sqrt{2\xi} $ and $\dot \gamma_0 \gg \sqrt{\lambda} M_{\rm pl} / \sqrt{2\xi} $, we may solve Eq. (\ref{gammasol}) and recover the forms of $\gamma(t)$ presented in Eqs. (\ref{dotgammalinear}) and (\ref{dotgammanonlinear}).

\acknowledgements{It is a pleasure to thank Alan Guth, Mustafa Amin, and Leo Stein for helpful discussions. This work was supported in part by the U.S. Department of Energy (DoE) under contract No. DE-FG02-05ER41360.}


\begin{thebibliography}{999}
\bibitem{Footnote1} Because we are working only to first order in perturbations, we define $Q^I$ in Eq. (\ref{Q}) in terms of $\delta \phi^I$ rather than the field-space vector ${\cal Q}^I$ that is introduced in \cite{KMSBispectrum}, since, as shown there, $\delta \phi^I \rightarrow {\cal Q}^I + {\cal O} ({\cal Q}^2 )$.
\end{thebibliography}
\end{document}